\documentclass[12pt]{article}
\usepackage{graphicx}
\usepackage{amssymb}
\usepackage{amsmath}
\usepackage{mciteplus}

\def\pavia{INFN - Sezione di Pavia, I-27100 Pavia, ITALY}
\def\Title#1{\begin{center} {\Large #1 } \end{center}}
\def\Author#1{\begin{center}{ \sc #1} \end{center}}
\def\Address#1{\begin{center}{ \it #1} \end{center}}

\newenvironment{Abstract}{\begin{quotation}  }{\end{quotation}}
\newenvironment{Presented}{\begin{quotation} \begin{center} 
             PRESENTED AT\end{center}\bigskip 
      \begin{center}\begin{large}}{\end{large}\end{center} \end{quotation}}
\def\Acknowledgements{\bigskip  \bigskip \begin{center} \begin{large}
             \bf ACKNOWLEDGEMENTS \end{large}\end{center}}
\begin{document}
\begin{titlepage}
%\pubblock

\vfill
\Title{First extraction of transversity from data \\ on lepton-hadron scattering and hadronic collisions}
\vfill
%\Author{ Despina Reggiano\support}
\Author{ Marco Radici}
\Address{\pavia}
\vfill
\begin{Abstract}
We present the first extraction of the transversity distribution based on the global analysis of pion-pair production in deep-inelastic scattering and in proton-proton collisions with one transversely polarized proton. The extraction relies on the knowledge of di-hadron fragmentation functions, which are taken from the analysis of electron-positron annihilation data. For the first time, the chiral-odd transversity is extracted from a global analysis similar to what is usually done for the chiral-even spin-averaged and helicity distributions. The knowledge of transversity is important among other things for detecting possible signals of new physics in high-precision low-energy experiments. 
\end{Abstract}
\vfill
\begin{Presented}
13$^{\mathrm{th}}$ Conference on the Intersections of \\ Particle and Nuclear Physics (CIPANP 2018) \\
Palm Springs, California (USA),  May 29 -- June 3, 2018
\end{Presented}
\vfill
\end{titlepage}
\def\thefootnote{\fnsymbol{footnote}}
\setcounter{footnote}{0}

%%%%%%%%%%%%%%%%%%%%%%%%%%%%%%

\section{Introduction}
\label{sec:intro}

The transversely polarized parton distribution $h_1$ (transversity) is the least known of parton distribution functions (PDFs) because it is not diagonal in the helicity basis (in jargon, it is a chiral-odd function). As such, $h_1$ is suppressed in simple processes like inclusive deep-inelastic scattering (DIS). It can be measured only in processes with at least two hadrons: semi-inclusive DIS (SIDIS) or hadronic collisions. On the other hand, transversity has recently received increasing attention because its first Mellin moment, the so-called tensor charge $\delta q$ (for a flavor $q$), can be useful to search for effects from physics beyond the Standard Model (BSM)~\cite{Courtoy:2015haa}. 

%In general, the knowledge of PDFs is crucial for the interpretation of high-energy experiments involving hadrons and for detecting signals of new physics beyond the Standard Model (BSM). Transversity has recently received increasing attention because of the importance of a precise  determination of its integral, 
%\begin{equation}
%\delta q (Q^2) = \int_0^1 dx \, h_1^{q - \bar{q}} (x, Q^2) \; ,
%\label{eq:tensorcharge}
%\end{equation}
%the so-called tensor charge $\delta q$ (for a flavor $q$)~\cite{Courtoy:2015haa}. In neutron $\beta$-decays, BSM effects can arise from the interference between SM operators and a new possible tensor operator whose coupling involves $\delta q$~\cite{Bhattacharya:2011qm}. Similarly, $\delta q$ enters the expression of the fermionic electric dipole moment that could constrain new possible CP-violating couplings in some BSM theories~\cite{Dubbers:2011ns,Yamanaka:2017mef}. 

In this work, we want to access the transversity PDF at leading twist and in the standard framework of collinear factorization. We consider the semi-inclusive production of two hadrons with small invariant mass, where there is a correlation between the transverse polarization of the quark directly fragmenting into the two hadrons and their transverse relative momentum~\cite{Collins:1994ax}. In this case, the di-hadron SIDIS cross section (once integrated over partonic transverse momenta) contains a specific modulation in the azimuthal orientation of the plane containing the momenta of the two hadrons. The coefficient of this modulation is the simple product $h_1 H_1^\sphericalangle$ where $H_1^\sphericalangle$ is a chiral-odd di-hadron fragmentation function (DiFF) quantifying the above correlation~\cite{Bianconi:1999cd,Radici:2001na,Bacchetta:2002ux}. 
The function $H_1^\sphericalangle$ can be independently determined by looking at correlations between the azimuthal orientations of two hadron pairs in back-to-back jets in $e^+ e^-$ 
%annihilation~\cite{Boer:2003ya,Bacchetta:2008wb,Courtoy:2012ry,Matevosyan:2018icf}. 
annihilation~\cite{Boer:2003ya,Courtoy:2012ry,Matevosyan:2018icf}. 
The advantage of this method is that collinear factorization makes it possible to isolate the same combination $h_1 H_1^\sphericalangle$ also in proton-proton collisions~\cite{Bacchetta:2004it}, giving rise to an azimuthally asymmetric distribution of the final hadron pair when one of the two initial protons is transversely 
%polarized~\cite{Radici:2016lam}. 
polarized~\cite{Radici:2016lam}. 

Experimental data for the SIDIS asymmetry in the azimuthal distribution of final $(\pi^+ \pi^-)$ pairs were collected by the {\tt HERMES} collaboration for a proton target~\cite{Airapetian:2008sk}, and by the {\tt COMPASS} collaboration for both protons and deuterons~\cite{Adolph:2012nw,Adolph:2014fjw}. The azimuthal asymmetry in the distribution of back-to-back $(\pi^+ \pi^-)$ pairs in $e^+ e^-$ annihilation was measured by the {\tt BELLE} collaboration~\cite{Vossen:2011fk}, opening the way to the first parametrization of $H_1^\sphericalangle$~\cite{Courtoy:2012ry}. In turn, this result was used in combination with the SIDIS data to extract the valence components of 
%$h_1$~\cite{Bacchetta:2011ip,Bacchetta:2012ty,Radici:2015mwa}. 
$h_1$~\cite{Bacchetta:2012ty,Radici:2015mwa}. 
Recently, the {\tt STAR} collaboration released the first results for the predicted asymmetry in the azimuthal distribution of $(\pi^+ \pi^-)$ pairs produced in proton-proton collisions with a transversely polarized proton~\cite{Adamczyk:2015hri}. 
%~\footnote{Recently, the {\tt STAR} collaboration has released new data at higher center-of-mass energy $\sqrt{s} = 500$ GeV~\cite{Adamczyk:2017ynk}. We will include these data in a future update of our work.}. 
Here, we discuss the extraction of the transversity PDF $h_1$ from a global fit of all these data for the semi-inclusive production of $(\pi^+ \pi^-)$ pairs. The final results have been published in Ref.~\cite{Radici:2018iag}.

%%%%%%%%%%%%%%%%%%%%%%%%%%%%%%

\section{Formalism}
\label{sec:formulae}

The relevant observables for our global fit are the following. 

For SIDIS on a transversely polarized hadron target, at leading twist the cross section contains the azimuthally asymmetric term $\sin (\phi^{}_S + \phi^{}_R) \, h_1 \, H_1^{\sphericalangle}$, where $\phi^{}_S$ and $\phi^{}_R$ are the azimuthal angles with respect to the scattering plane of the target polarization vector and pair relative momentum, respectively. The corresponding single-spin asymmetry is described in detail in 
%Refs.~\cite{Bacchetta:2011ip,Bacchetta:2012ty, Radici:2015mwa}. 
Refs.~\cite{Bacchetta:2012ty, Radici:2015mwa}. 

The $H_1^{\sphericalangle}$ can be extracted from the electron-positron annihilation process leading to two correlated hadron pairs in opposite hemispheres. Again at leading twist, the cross section contains the azimuthally asymmetric term $\cos (\phi^{}_R + \bar{\phi}^{}_R) \, H_1^{\sphericalangle} \bar{H}_1^{\sphericalangle}$, where the barred quantities refer to the corresponding antiquark emerging from the $e^+ e^-$ annihilation. From this, we can construct the so-called Artru-Collins asymmetry~\cite{Boer:2003ya,Courtoy:2012ry}, and extract the $H_1^{\sphericalangle}$ by summing upon all pairs in one hemisphere, involving the number density of fragmenting (un)polarized quarks~\cite{Courtoy:2012ry}. 

Lastly, at leading order in the hard scale for proton-proton collisions with one transversely polarized proton the cross section contains the  azimuthally asymmetric term $\sin (\phi^{}_S - \phi^{}_R) \, f_1^a \otimes h_1^b \otimes d\sigma_{ab^\uparrow \to c^\uparrow d} \otimes H_1^{\sphericalangle\, c}$, where $\phi^{}_S, \, \phi^{}_R$ have the same meaning as before but with respect to the reaction plane of the collision process, and $d\sigma$ is the elementary cross section for the annihilation of a parton $a$ with a transversely polarized parton $b$ into the parton $d$ and the transversely polarized parton $c$ (summing over all undetected $d$). This term leads to the spin asymmetry described in 
%Refs.~\cite{Bacchetta:2004it,Radici:2016lam}. 
Ref.~\cite{Bacchetta:2004it}. 
Despite working in the collinear framework, the elementary combination $h_1 \, H_1^{\sphericalangle}$ happens convoluted with other ingredients in the numerator of such asymmetry, making its repeated computation very demanding for a global fit. In order to speed up the execution of the fitting code, the well known workaround is to rewrite the parametric part of the integrand (in our case, the transversity) in terms of its Mellin anti-transform, such that most of the integrals can be computed before the minimization. In order to exploit this workaround, it is crucial that the Mellin transform of $h_1$ can be analytically calculated at any scale. 

To this purpose, we have chosen the following functional form for the valence component $q_v$ at the starting scale $Q_0^2 = 1$ GeV$^2$:
\begin{eqnarray}
& &\hspace{-1cm} x\, h_1^{q_v} (x, Q_0^2) = F^q (x) \, F_{\mathrm{SB}}^q (x) \, , \label{e:h1xQ0} \\
& &\hspace{-1cm} F_{\mathrm{SB}}^q (x) = N_{\mathrm{SB}}^q \, x^{a_q} \, (1-x)^{b_q} \, \left( 1+ c_q \, \sqrt{x} + d_q \, x + e_q \, x^2 + f_q \, x^3 \right) \, , \label{e:SBfit} \\
& &\hspace{-1cm} F^q (x) = N_F^q \, \frac{{\cal F}^q (x)}{\mathrm{max}_x [|{\cal F}^q (x)|]} \, , \quad 
{\cal F}^q (x) = x^{A_q} \, \left[ 1 + B_q \, T_1(x) + C_q \, T_2 (x) + D_q \, T_3 (x) \right]  .\label{e:Fpar}
\end{eqnarray}
The $F_{\mathrm{SB}}$ in Eq.~(\ref{e:SBfit}) is a fit to the Soffer bound at $Q_0^2$, whose analytic expression is listed in the appendix of Ref.~\cite{Bacchetta:2012ty}. Using the parameters listed in Tab.I of Ref.~\cite{Radici:2018iag}, the accuracy is of order 1\% in the range $0.001\leq x \leq 1$. In Eq.~(\ref{e:Fpar}), the $T_n (x)$ are the Cebyshev polynomials of order $n$. The $N_F^q, \, A_q, \, B_q, \, C_q, \, D_q, $ are the fitting parameters. 
At leading twist, we can make simplifying assumptions on DiFFs with respect to isospin symmetry and charge conjugation such that only the valence components of transversity can be 
%accessed~\cite{Bacchetta:2006un,Bacchetta:2011ip}. 
accessed~\cite{Bacchetta:2006un}. 
Hence, we have in total 10 free parameters. If we impose the constraint $|N_F^q | \leq 1$, then $|F^q (x)| \leq 1$ for all $x$, and the $h_1^{q_v}$ in Eq.~(\ref{e:h1xQ0}) automatically satisfies the Soffer inequality at any $x$ and at any scale. Another constraint comes from the low-$x$ behavior of transversity, which from the above equations looks like $x h_1^{q_v}(x) \approx x^{A_q + a_q}$. If we obviously demand that the tensor charge $\delta q (Q^2)$ is finite, this translates into the constraint $A_q + a_q > 0$. According to Ref.~\cite{Accardi:2017pmi}, the more stringent condition $A_q + a_q > 1$ is required to avoid a violation of the Burkardt-Cottingham sum rule by an infinite amount. Recently, Kovchegov and Sievert calculated the transversely polarized dipole scattering amplitude in the large-$N_c$ limit~\cite{Kovchegov:2018zeq} (with $N_c$ the number of colors) and obtained that at very small $x$ the transversity behaves as if $A_q + a_q \approx 1$. As in Ref.~\cite{Radici:2018iag}, here we numerically evaluate the tensor charge in the range $[10^{-6}, 1]$. In order to avoid uncontrolled extrapolation errors at $0\leq x < 10^{-6}$, we impose the condition $A_q + a_q > 1/3$, which also grants that $\delta q$ is evaluated at 1\% accuracy.

The statistical uncertainty of the global fit is studied at the 90\% confidence level using the same bootstrap method as in our previous fits~\cite{Bacchetta:2012ty,Radici:2015mwa}. In the analysis of di-hadron $e^+ e^-$ data, the unconstrained gluon channel is assumed $D_1^g (Q_0^2) = 0$. We parametrize this error by computing the asymmetry also  with $D_1^g Q_0^2) = D_1^u (Q_0^2) / 4$, or $D_1^u (Q_0^2)$. We have verified that these choices alter the $\chi^2$ of the $e^+ e^-$ fit in Ref.~\cite{Courtoy:2012ry} by 10-50\%, keeping always $\chi^2$/d.o.f. $\lesssim 2$ . The number $M$ of replicas is fixed by reproducing the mean and standard deviation of the original data points. For each option, it turns out that 200 replicas are sufficient. Hence, we have in total $M=600$ replicas. The {\tt STAR} data are presented as three projections in different bins of the same data set. We use all the projections but we multiply the statistical errors by the square root of their corresponding weight with respect to the total number of bins; at the same time, when computing the actual degrees of freedom we reduce the number of bins by the same coefficients. With 10 free parameters, we get a final $\chi^2$ per degree of freedom equal to $1.76 \pm 0.11$, where the $\chi^2$ of all the replicas are distributed approximately as a Gaussian.

%%%%%%%%%%%%%%%%%%%%%%%%%%%%

\section{Results}
\label{sec:outcome}

%%%%%%%%%% Fig. 1  %%%%%%%%%
\begin{figure}
\begin{center}
\includegraphics[width=0.4\textwidth]{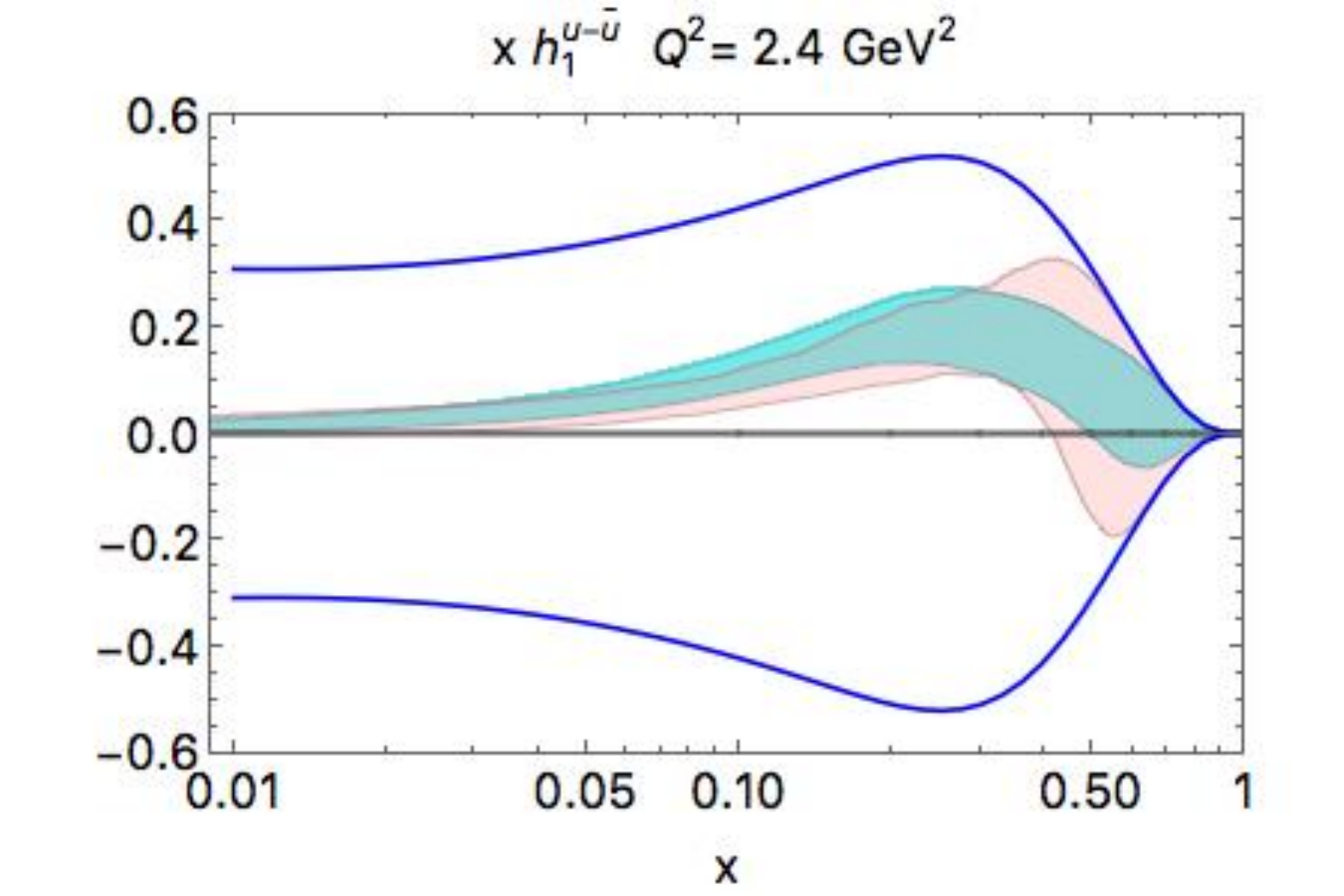} \hspace{0.1cm} 
\includegraphics[width=0.4\textwidth]{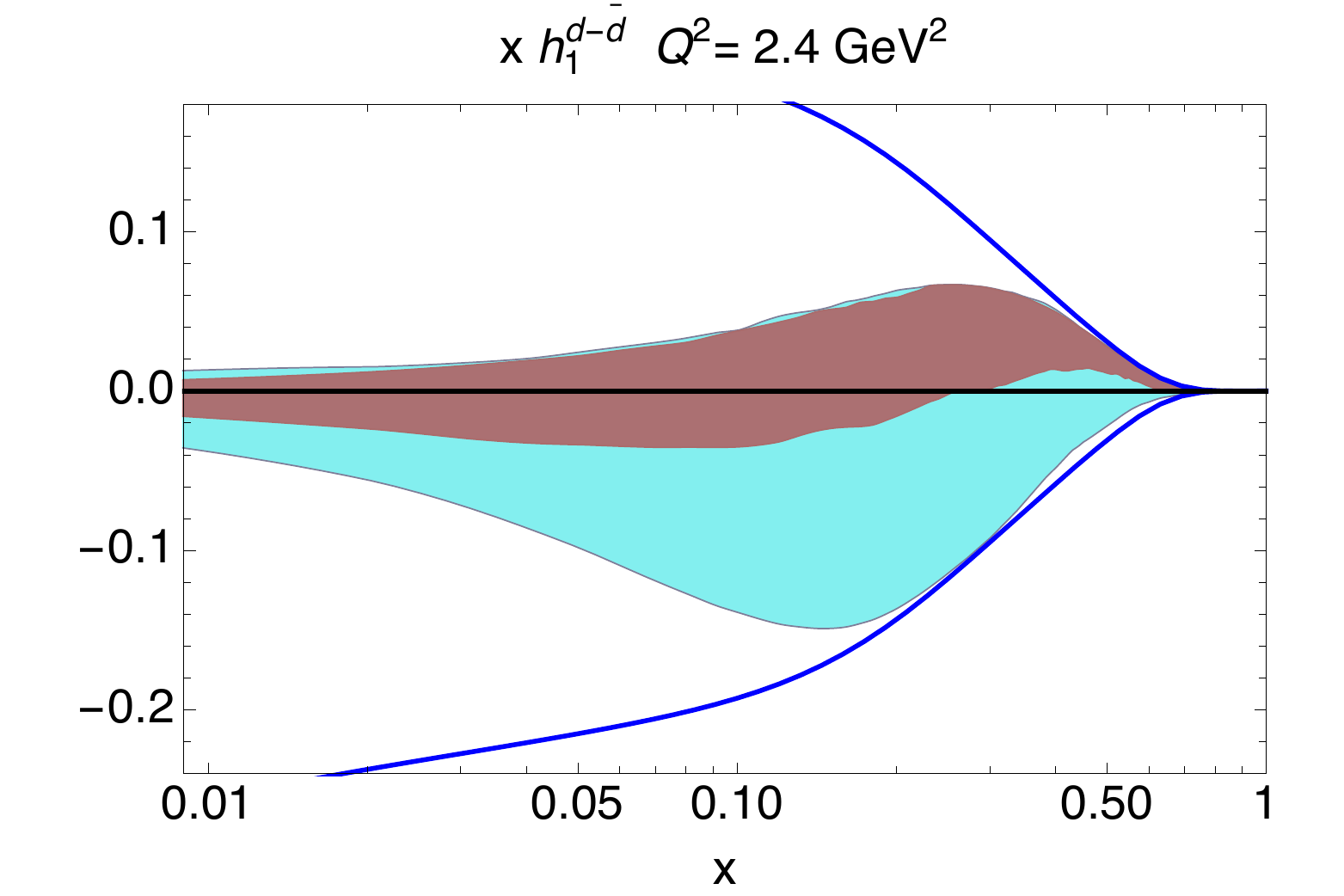}
\end{center}
\caption{The transversity $x\, h_1$ as a function of $x$ at $Q^2 = 2.4$ GeV$^2$. Blue lines represent the Soffer bounds. Cyan bands for the global fit of this work including all options $D_1^g (Q_0^2) = 0$, $D_1^u (Q_0^2) / 4$ and $D_1^u (Q_0^2)$. Left panel for valence up quark: comparison with our previous fit in Ref.~\cite{Radici:2015mwa} (pink band). Right panel for valence down quark: comparison with this global fit with only $D_1^g (Q_0^2) = 0$ (darker pink band in the foreground).}
\label{fig:oldnewfit}
\end{figure}
%%%%%

In Fig.~\ref{fig:oldnewfit}, the transversity $x\, h_1$ is displayed as a function of $x$ at $Q^2 = 2.4$ GeV$^2$. The blue lines represent the Soffer bounds. The left panel refers to the valence up component. Here, the pink band corresponds to our previous fit with only SIDIS and $e^+ e^-$ data~\cite{Radici:2015mwa}. The cyan band is the new global fit discussed here, including all options $D_1^g (Q_0^2) = 0$, $D_1^g (Q_0^2) = D_1^u (Q_0^2) / 4$ and $D_1^g (Q_0^2) = D_1^u (Q_0^2)$. It turns out that this result is insensitive to the various choices for $D_1^g (Q_0^2)$~\cite{Radici:2018iag}. There is an evident gain in precision by including also the {\tt STAR} $p$-$p$ data. The uncertainty of our previous fit in Ref.~\cite{Radici:2015mwa} (the pink band) is comparable to the one obtained from the analysis of the Collins effect~\cite{Kang:2015msa,Anselmino:2013vqa}. Hence, we deduce that the outcome of our global fit provides a substantial increase in the precision on $h_1^{u_v}$ and on the related tensor charge $\delta u$ with respect to all the other phenomenological extractions. 

The right panel of Fig.~\ref{fig:oldnewfit} refers to the valence down component. Again, the cyan band corresponds to the new global fit including all options for $D_1^g (Q_0^2)$. The darker pink band in the foreground shows how the result is modified by including only the option $D_1^g (Q_0^2) = 0$. Evidently, the valence down component $h_1^{d_v}$ is very sensitive to $D_1^g$. In the cross section for proton-proton collisions, the up and down quarks have the same weight and gluon contributions occur already at leading order (LO). Hence, we would expect that both components of transversity are sensitive to the various choices for $D_1^g (Q_0^2)$. However, in the SIDIS cross section the gluon contributions appear only at NLO and are here neglected. Moreover, the up quark component is emphasized by a factor 8; the net result is that it is rather insensitive to the various choices for $D_1^g (Q_0^2)$~\cite{Radici:2018iag}. In this respect, data on $(\pi^+ \pi^-)$ multiplicities in $p$-$p$ collisions would be very useful. Finally, we notice that the unnatural behavior of $h_1^{d_v}$ at $x \gtrsim 0.1$ obtained in Ref.~\cite{Radici:2015mwa} has disappeared. Hence, the {\tt STAR} data have a significant impact on our knowledge of transversity.

%%%%%%%%%% Fig. 2  %%%%%%%%%
\begin{figure}
\begin{center}
\includegraphics[width=0.37\textwidth]{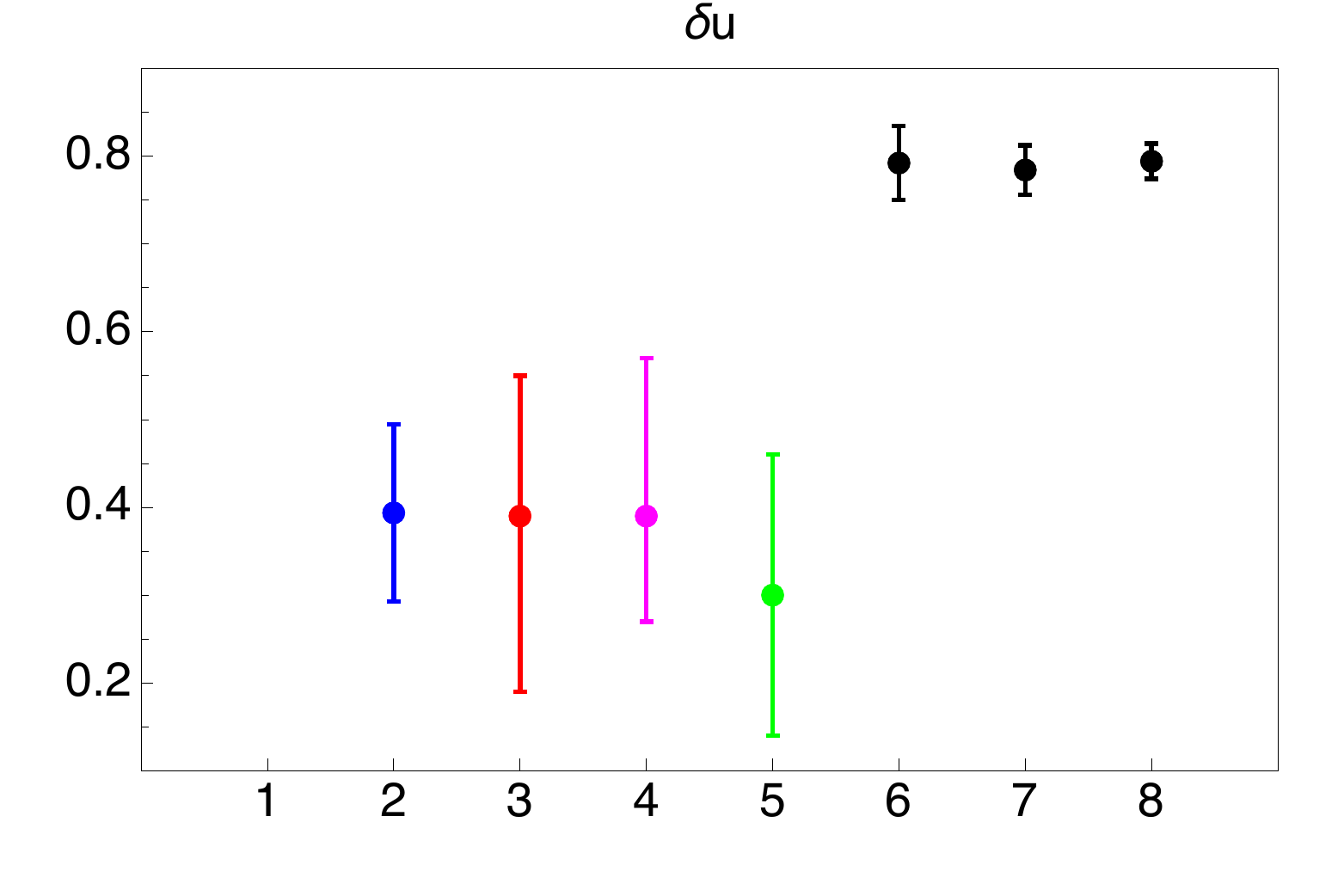} \hspace{0.1cm} 
\includegraphics[width=0.37\textwidth]{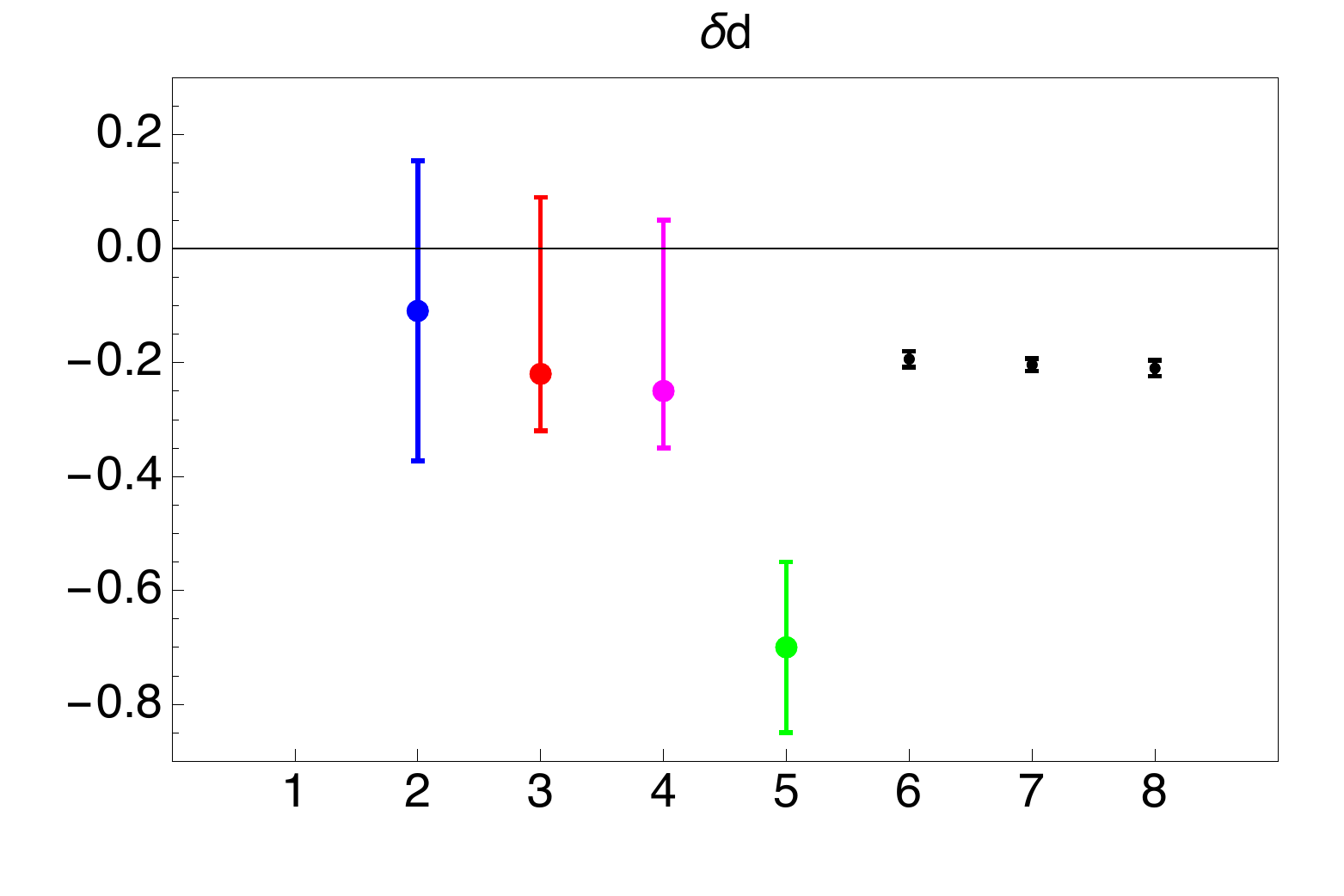} %\\
\end{center}
\caption{The tensor charges at $Q^2 = 4$ GeV$^2$ and 90\% confidence level. Blue point (label 2) for this global fit, red point (label 3) from Ref.~\cite{Kang:2015msa}, magenta point (label 4) from Ref.~\cite{Anselmino:2013vqa}, green point (label 5) from Ref.~\cite{Lin:2017stx}, black points (labels 6, 7 and 8) from Ref.~\cite{Bhattacharya:2016zcn,Gupta:2018lvp,Alexandrou:2017qyt}, respectively. Left panel for valence up quark, right panel for valence down quark.}
%\caption{Upper panels: the tensor charges $\delta u$ (left) and $\delta d$ (right) at $Q^2 = 4$ GeV$^2$ and 90\% confidence level; blue point (label 2) for this global fit, red point (label 3) from Ref.~\cite{Kang:2015msa}, magenta point (label 4) from Ref.~\cite{Anselmino:2013vqa}, green point (label 5) from Ref.~\cite{Lin:2017stx}, black points (labels 6, 7 and 8) from Ref.~\cite{Bhattacharya:2016zcn,Gupta:2018lvp,Alexandrou:2017qyt}, respectively. Lower panel: the isovector tensor charge $g_T = \delta u - \delta d$ with same notations and color codes; additional lattice calculations in black points (labels 9, 10 and 11) from Refs.~\cite{Chang:2017eiq,Green:2012ej,Bali:2014nma}, respectively.}
\label{fig:tensorcharge}
\end{figure}
%\vspace{-0.1cm}
%%%%%

In Fig.~\ref{fig:tensorcharge}, we show the tensor charge $\delta q$ at $Q^2 = 4$ GeV$^2$ and 90\% confidence level. The left panel refers to the up quark $\delta u$, the right panel to the down quark $\delta d$%, and the lower panel to the isovector component $g_T = \delta u - \delta d$. 
In all panels, the blue point with label 2 is the result of this global fit, the red point with label 3 is the result of the phenomenological extraction based on the Collins effect in the transverse-momentum dependent framework~\cite{Kang:2015msa}, the magenta point with label 4 is also based on the Collins effect but in an extended parton model framework~\cite{Anselmino:2013vqa}, the green point with label 5 is the result obtained with a nested Monte Carlo iterative approach to the Collins effect but with the additional constraint to reproduce the lattice results for $g_T$~\cite{Lin:2017stx}, the black points with labels 6, 7 and 8, are the latest lattice results from the PNDME~\cite{Bhattacharya:2016zcn,Gupta:2018lvp} and ETMC~\cite{Alexandrou:2017qyt} collaborations, respectively. 
%In the lower panel, the black points with labels 9, 10 and 11, refer to Refs.~\cite{Chang:2017eiq,Green:2012ej,Bali:2014nma}, respectively. 
Our results seem in very good agreement with other phenomenological extractions except for Ref.~\cite{Lin:2017stx} on $\delta d$ (and $g_T$). 

In general, there seems to be compatibility between lattice simulations and phenomenology only for $\delta d$ but not for $\delta u$ (and, consequently, for $g_T$). In particular, it seems that for $\delta u$ the more precise the phenomenological extractions are the larger the deviation is from lattice.  But before claiming that there is a "transverse spin puzzle", more work is needed along both lines of improving the precision of phenomenological extractions and of benchmarking lattice simulations. In this perspective, in the following section we check the impact of very precise SIDIS pseudodata on a deuteron target from {\tt COMPASS}.

%%%%%%%%%%%%%%%%%%%%%%%%%%%%

\section{Impact of {\tt COMPASS} deuteron pseudodata}
\label{sec:compass}

In the following, we discuss the impact on our global fit of including also a new set of SIDIS pseudodata on deuteron targets from the {\tt COMPASS} collaboration, that reflect the precision that will be reached in the scheduled 2021 run. The pseudodata are taken at the same bins of the 2004 run~\cite{Adolph:2012nw}, but the higher precision corresponds to a statistical error reduced to approximately 60\% of the statistical error of the 2010 run on proton targets~\cite{Adolph:2014fjw}. The arbitrary average value of each data point is conveniently chosen to coincide with the average value of the 90\% uncertainty band of replicas from previous global fit computed at the same bin. 

%%%%%%%%%% Fig. 3  %%%%%%%%%
\begin{figure}
\begin{center}
\includegraphics[width=0.4\textwidth]{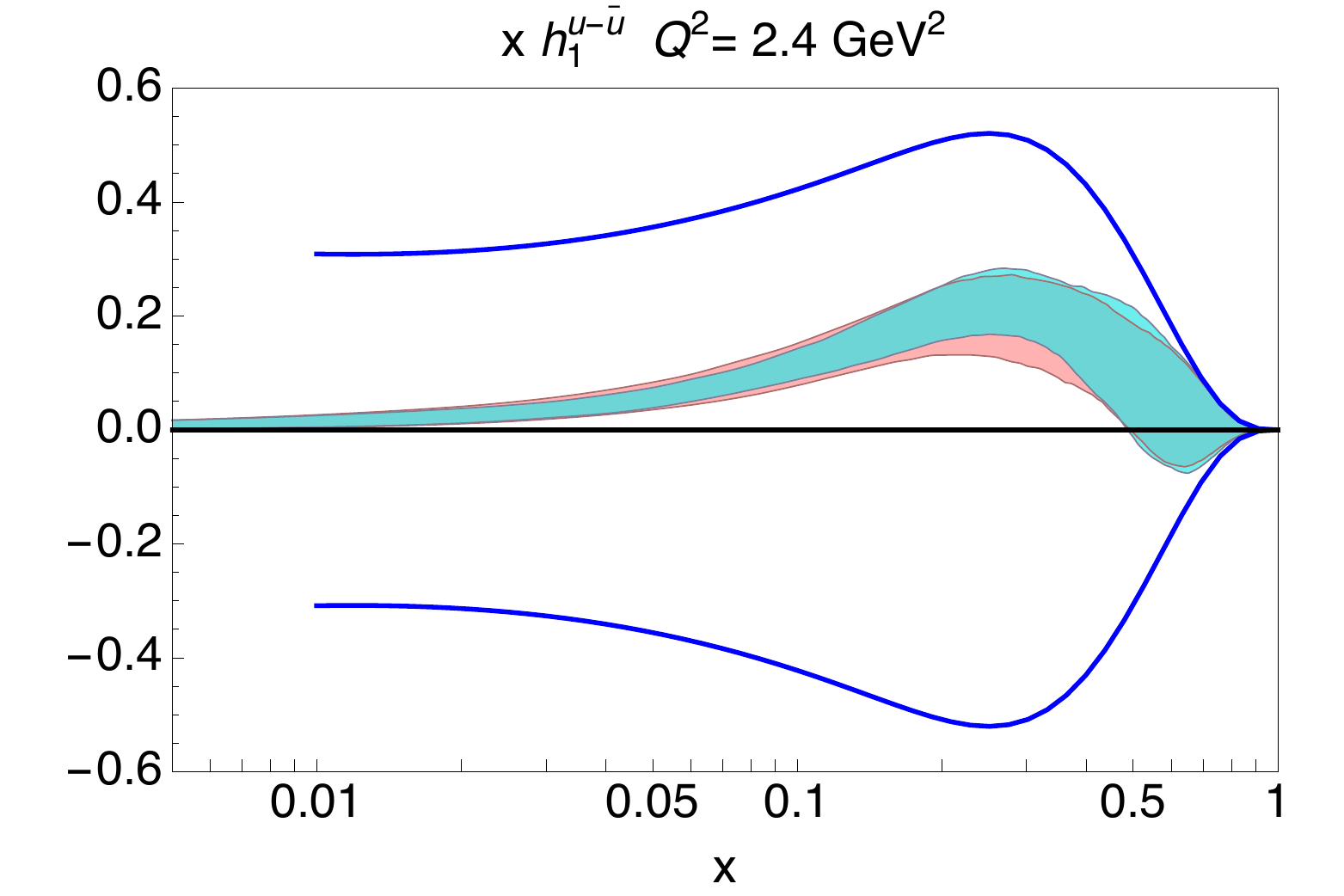} \hspace{0.1cm} 
\includegraphics[width=0.4\textwidth]{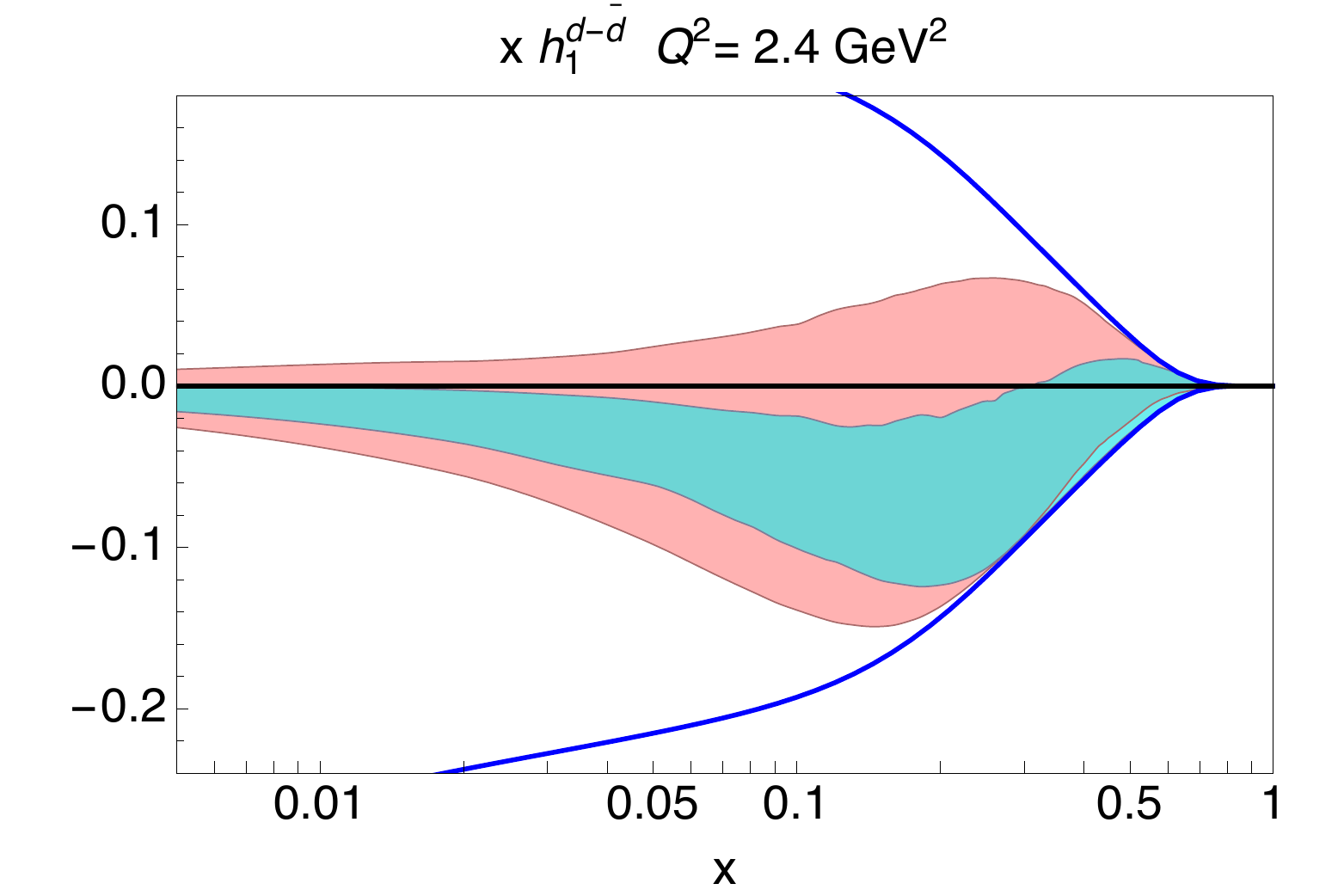}
\end{center}
\caption{The transversity $x\, h_1$ as a function of $x$ at $Q^2 = 2.4$ GeV$^2$. Blue lines represent the Soffer bounds. Left panel for valence up quark, right panel for valence down quark. In both panels, the pink band in the background corresponds to the cyan band in Fig.~\ref{fig:oldnewfit}. Here, the cyan band is the result of the global fit including the {\tt COMPASS} pseudodata.}
\label{fig:pseudofit}
\end{figure}
%%%%%

In Fig.~\ref{fig:pseudofit}, the transversity $x\, h_1$ is shown again as a function of $x$ at $Q^2 = 2.4$ GeV$^2$. Blue lines represent the Soffer bounds. The left panel refers to the valence up quark, the right panel to the valence down quark. The pink bands in the background coincide with the cyan bands in Fig.~\ref{fig:oldnewfit}, namely they represent the results of the global fit discussed in the previous section when the maximum uncertainty is reached by including all options for $D_1^g (Q_0^2)$. Here, the cyan bands show the impact of including in the fit also the SIDIS pseudodata on deuteron targets from the {\tt COMPASS} collaboration. For the down quark, the increase in precision is evident and amounts to reducing the width of the band by almost a factor 2 in the $x$-range covered by pseudodata. For the up quark, the effect is less evident. But if one plots the ratio of the cyan band to the pink one as a function of $x$, it is easy to realize that on average there is an almost 20\% increase in precision in the $x$-range of pseudodata. The global $\chi^2$ per degree of freedom is reduced to $1.32 \pm 0.09$. 

%%%%%%%%%% Fig. 4  %%%%%%%%%
\begin{figure}
\begin{center}
\includegraphics[width=0.37\textwidth]{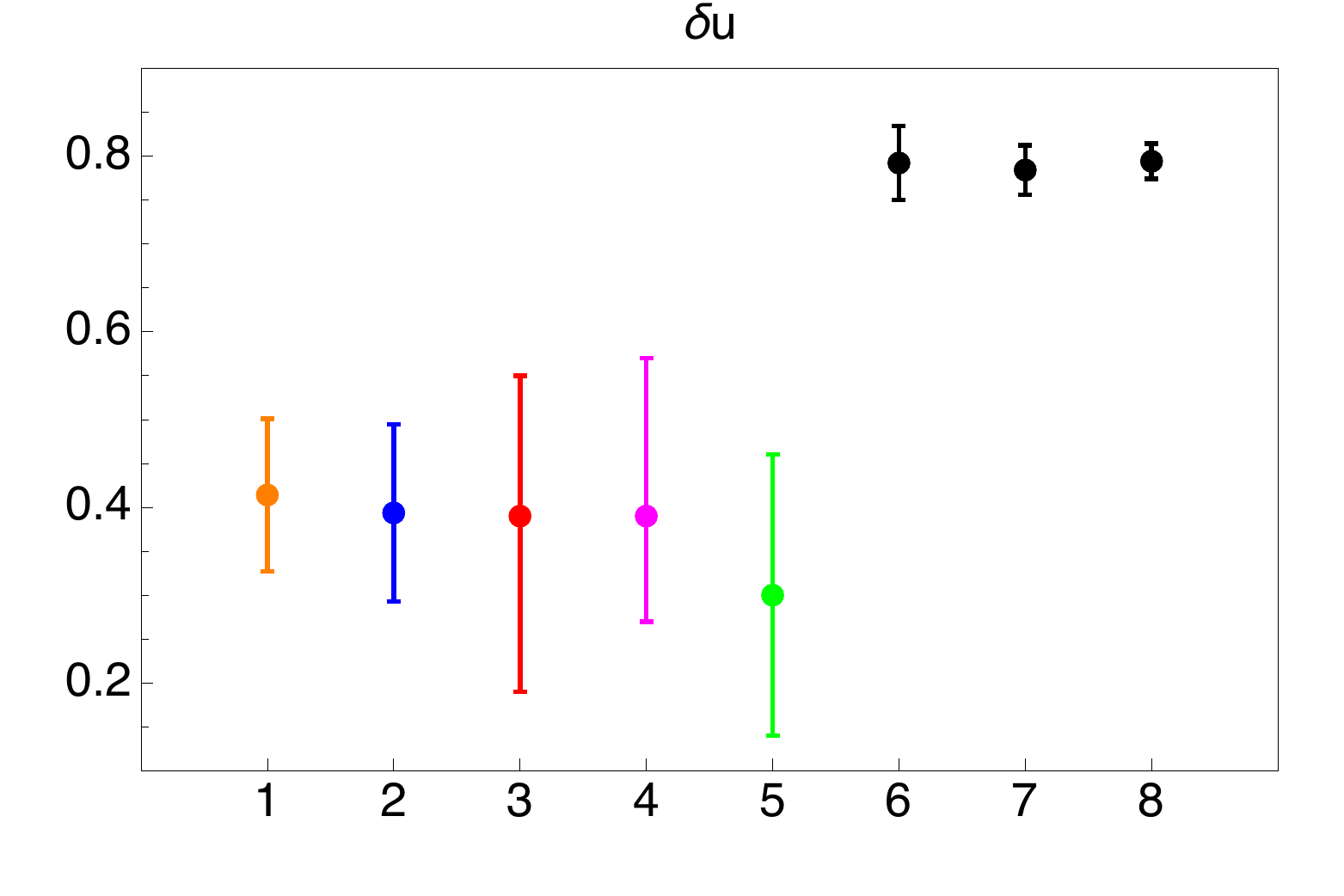} \hspace{0.1cm} 
\includegraphics[width=0.37\textwidth]{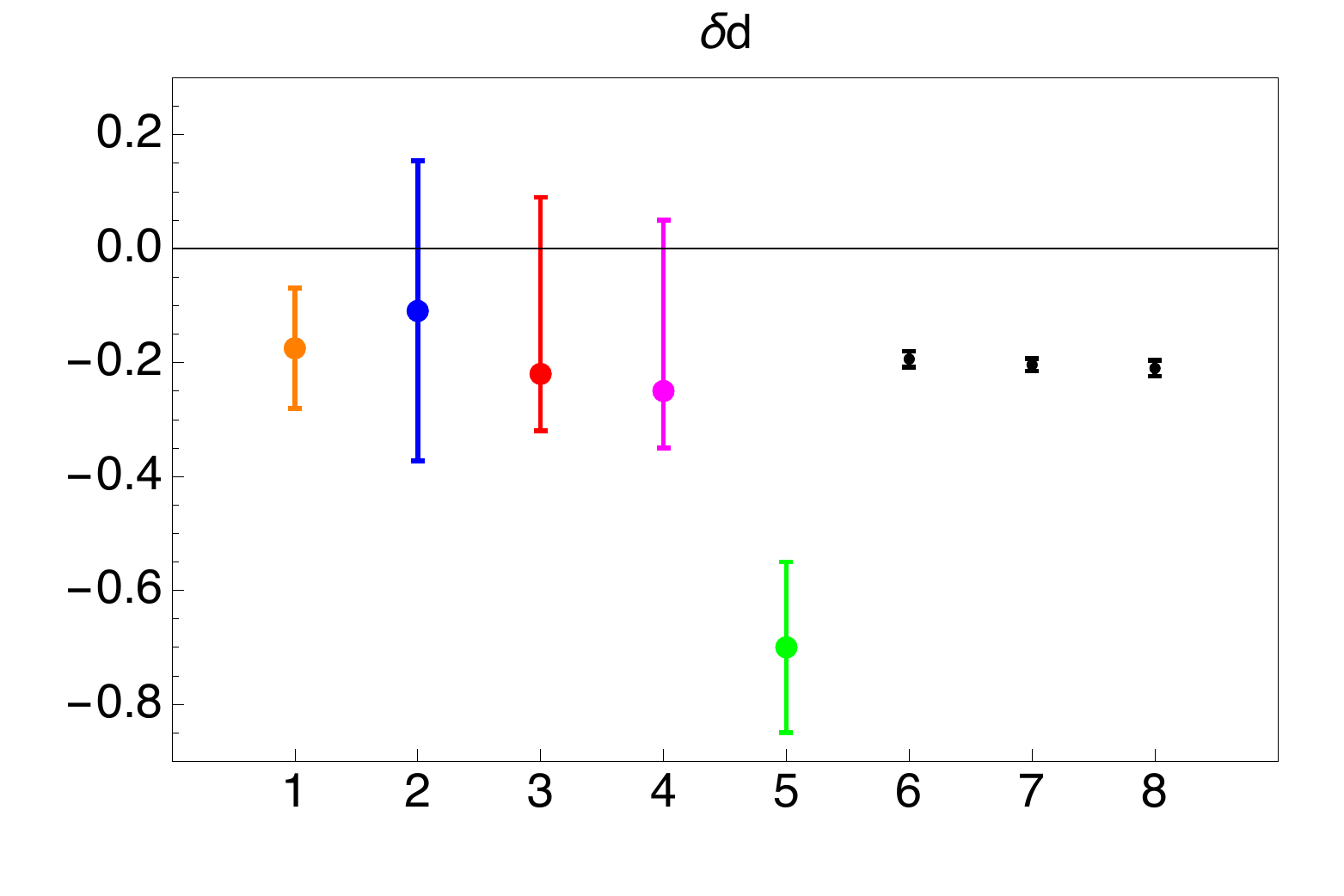} %\\
\end{center}
%\caption{Upper panels: the tensor charges $\delta u$ (left) and $\delta d$ (right). Lower panel: the isovector tensor charge $g_T = \delta u - \delta d$. All results at $Q^2 = 4$ GeV$^2$ and 90\% confidence level. Notations, color codes and conventions as in Fig.~\ref{fig:tensorcharge}. The orange point (label 1) is the result of the global fit when including the SIDIS deuteron pseudodata from {\tt COMPASS}.
\caption{The tensor charges $\delta u$ (left) and $\delta d$ (right) at $Q^2 = 4$ GeV$^2$ and 90\% confidence level. Notations, color codes and conventions as in Fig.~\ref{fig:tensorcharge}. The orange point (label 1) is the result of the global fit when including the SIDIS deuteron pseudodata from {\tt COMPASS}.}
\label{fig:pseudotensorcharge}
\end{figure}
%\vspace{-0.1cm}
%%%%%

In Fig.~\ref{fig:pseudotensorcharge}, we show the tensor charge in the same conditions, notations and conventions as in Fig.~\ref{fig:tensorcharge} except for the orange point with label 1, which corresponds to the result of the global fit when including the SIDIS deuteron pseudodata from {\tt COMPASS}. By comparing the orange and blue points, we can estimate the impact of these pseudodata on the global fit: the increase of precision is evident, particularly for the down quark. The orange points represent the currently most precise phenomenological estimate of the proton tensor charge. However, they confirm the general trend: there is an overall agreement among various phenomenological extractions of transversity (except for the down and isovector results of Ref.~\cite{Lin:2017stx}), but there is a trend towards a clear disagreement with lattice (except for the down tensor charge).

%%%%%%%%%%%%%%%%%%%%%%%

\section{Conclusions}
\label{sec:end}

In summary, we have presented a new phenomenological extraction of the transversity distribution in the framework of collinear factorization by performing for the first time a global fit of all data for the semi-inclusive production of charged pion pairs in deep--inelastic scattering, electron-positron annihilation, and proton-proton collisions. The uncertainty on the result for the valence up quark is smaller than any previous extraction. The large sensitivity of the valence down quark to the unconstrained gluon channel in di-hadron fragmentation calls for data on pion pair multiplicities, particularly in proton-proton collisions which are yet missing. The calculated (isovector) tensor charge seems compatible with most phenomenological extractions but not with lattice calculations. This trend is confirmed by including in the global fit very precise {\tt COMPASS} pseudodata for SIDIS on a transversely polarized deuteron target. 

%%%%%%%%%%%%%%%%%%%%%%%

\Acknowledgements

Most of the results presented in this report have been carried out in collaboration with A.~ Bacchetta, to whom I am deeply indebted. This research is partially supported by the European Research Council (ERC) under the European Union's Horizon 2020 research and innovation program (Grant Agreement No. 647981, 3DSPIN). 

%%%%%%%%%%%%%%%%%%%%%%%

\bibliographystyle{JHEP}
\bibliography{mybiblio}

%%%%%%%%%%%%%%%%%%%%%%%
 
\end{document}